\documentclass[prl,twocolumn,superscriptaddress,flushbottom,balancelastpage,a4paper,oneside]{revtex4}

\usepackage{latexsym,amsmath,amssymb,bm,graphicx}

\makeatletter\def\Dated@name{}\makeatother

\newcommand{\myappendixname}{appendix}
\newcommand{\mymhcite}{\cite{MuellerHartmann89}}

\newcommand{\mycaptionone}{%
  \textbf{Kinks in the dispersion relation $E_{\bm{k}}$ for 
    a strongly correlated system}.  The intensity plot represents
  the spectral function $A(\bm{k},\omega)$ (Hubbard model in DMFT, cubic lattice, interaction $U$=3.5 eV,
  bandwidth $W\approx$ 3.64 eV, $n$=1, $Z_{\text{FL}}$=0.086, $T=5$ K).  Close
  to the Fermi energy the effective dispersion (white dots) follows
  the renormalized band structure
  $E_{\bm{k}}=Z_{\text{FL}}\epsilon_{\bm{k}}$ (blue line).  For
  $|\omega|>\omega_\star$ the dispersion has the same shape but with a
  different renormalization,
  $E_{\bm{k}}=Z_{\text{CP}}\epsilon_{\bm{k}}-c\;\text{sgn}(E_{\bm{k}})$
  (pink line). Here $\omega_\star$=0.03 eV, $Z_{\text{CP}}=0.135$, and
  $c=0.018$ eV are all calculated (see \myappendixname) from
  $Z_{\text{FL}}$ and $\epsilon_{\bm{k}}$ (black line).  A subinterval
  of $\Gamma$-R (white frame) is plotted on the right, showing kinks
  at $\pm\omega_\star$ (arrows).%
}

\newcommand{\mycaptiontwo}{%
  \textbf{Local propagator and 
    self-energy for a strongly correlated system (parameters as Fig.~\ref{fig:strong}).}
  \textbf{a}, Correlation-induced three-peak spectral function
  $A(\omega)=-\text{Im}G(\omega)/\pi$ with dips at $\pm\Omega=0.45$
  eV.  \textbf{b}, Corresponding real part of the propagator,
  $-\text{Re}G(\omega)$, with minimum and maximum at
  $\pm\omega_{\text{max}}$ inside the central spectral peak.
  \textbf{c}, Real part of the self-energy with kinks at
  $\pm\omega_\star$ (blue circles), located at the points of maximum
  curvature of $\text{Re}G(\omega)$,
  ($\omega_\star=0.4\omega_{\text{max}}=0.03$ eV).  \textbf{d},
  $\omega-1/G(\omega)$ contributes to the self-energy. In general
  $\text{Re}[\omega-1/G(\omega)]$ (blue line) is linear in
  $|\omega|<\Omega$.  The other contribution to the self-energy is
  $-\Delta(G(\omega))\approx-(m_2-m_1^2)G(\omega)$ (to lowest order in
  the moments $m_i$ of $\epsilon_{\bm{k}}$; here $m_2-m_1^2$=0.5
  eV$^2$).  Therefore the nonlinearity of $-\text{Re}[G(\omega)]$ at
  $\pm\omega_\star$ determines the location of kinks.%
}

\newcommand{\myacknowledgements}{%
  We acknowledge discussions with V. I. Anisimov, R. Bulla, J. Fink, A. Fujimori, and D. Manske.
  This work was supported by Deutsche Forschungsgemeinschaft through
  Sonderforschungsbereiche 484 (K.B., M.K., D.V.) and 602 (T.P.) and
  the Emmy-Noether 
  program (K.H.), and in part by the Russian Basic Research foundation
  grants 05-02-16301, 05-02-17244, 06-02-90537 as well as by the RAS Programs
  \emph{Quantum macrophysics} and \emph{Strongly correlated electrons
    in semiconductors, metals, superconductors and magnetic
    materials}, Dynasty Foundation, Grant of President of Russia
  MK-2118.2005.02, interdisciplinary grant UB-SB RAS (I.N.).  We thank
  the John von Neumann Institute for Computing, Forschungszentrum
  J\"ulich, and the Norddeutsche Verbund f\"ur Hoch- und
  H\"ochstleistungsrechnen for computing time.
}

\newcommand{\mybibliography}{%
  
}

\begin{document}

  \title{Kinks in the dispersion of strongly correlated electrons}

  \author{K.\ Byczuk}

  \affiliation{Theoretical Physics III, Center for Electronic
    Correlations and Magnetism, Institute for Physics, University of
    Augsburg, 86135 Augsburg, Germany}

  \affiliation{Institute of Theoretical Physics, Warsaw University,
    ul.\ Ho\.za 69, PL-00-681 Warszawa, Poland}

  \author{M.\ Kollar}

  \affiliation{Theoretical Physics III, Center for Electronic
    Correlations and Magnetism, Institute for Physics, University of
    Augsburg, 86135 Augsburg, Germany}

  \author{K.\ Held}

  \author{Y.-F.\ Yang}

  \affiliation{Max-Planck Institute for Solid State Research,
    Heisenbergstr.~1, 70569 Stuttgart, Germany}

  \author{I.~A.\ Nekrasov}

  \affiliation{Institute for Electrophysics, Russian Academy of
    Sciences, Ekaterinburg, 620016, Russia}

  \author{Th.\ Pruschke}

  \affiliation{Institute for Theoretical Physics, University of
    G\"ottingen, Friedrich-Hund-Platz 1, 37077 G\"ottingen, Germany}

  \author{D.\ Vollhardt}

  \affiliation{Theoretical Physics III, Center for Electronic
    Correlations and Magnetism, Institute for Physics, University of
    Augsburg, 86135 Augsburg, Germany}

  \date{22-Sep-2006}

  \begin{widetext}\maketitle\end{widetext}

  \begin{bf}
    The properties of condensed matter are determined by
    single-particle and collective excitations and their interactions.
    These quantum-mechanical excitations are characterized by an energy $E$ and a
    momentum $\hbar\bm{k}$ which are related through their
    \emph{dispersion} $E_{\bm{k}}$.  The coupling of two excitations
    may lead to abrupt changes (\emph{kinks}) in the slope of the
    dispersion. Such kinks thus carry important information about
    interactions in a many-body system.
    For example, kinks detected at 40-70 meV below the Fermi level in
    the electronic 
    dispersion of high-temperature superconductors are taken as
    evidence for phonon \cite{Lanzara2002,Shen2002} or
    spin-fluctuation based \cite{He2001,Hwang2004} pairing mechanisms.
    Kinks in the electronic dispersion at binding energies ranging from 30 to
    800 meV are also found in various other metals
    \cite{Hengsberger1999,Valla1999,Rotenberg2000,Higashiguchi2005,Schafer2004,Menzel2005,Yang2005,Aiura2004,Iwasawa2005,Sun2005,Ronning2003,Yoshida2005,Graf2006}
    posing questions about their origins.
    Here we report a novel, purely electronic mechanism yielding kinks
    in the electron dispersions.
    It applies to strongly correlated metals whose spectral function
    shows well separated Hubbard subbands and central peak as, for
    example, in transition metal-oxides.  The position of the kinks and
    the energy range of validity of Fermi-liquid (FL) theory is
    determined solely by the FL renormalization factor and the bare,
    uncorrelated band structure.
    Angle-resolved photoemission spectroscopy (ARPES) experiments at
    binding energies outside the FL regime 
    can thus provide new, previously unexpected
    information about strongly correlated electronic systems.
  \end{bf}

  In systems with a strong electron-phonon coupling kinks in the
  electronic dispersion at 40--60 meV are well known \cite{Hengsberger1999,Valla1999,Rotenberg2000}.
  Collective excitations other than phonons, or
  even an altogether different mechanism, may be the origin of kinks
  detected at 40 meV in the dispersion of surface states of Ni(110) \cite{Higashiguchi2005}.  Surface
  states of ferromagnetic Fe(110) show similar kinks at 100-200 meV
  \cite{Schafer2004}, and even at 300 meV in Pt(110) -- far beyond any
  phononic energy scale \cite{Menzel2005}.   Kinks at unusually high
  energies are also found in transition-metal oxides
  \cite{Ronning2003,Yang2005,Aiura2004,Iwasawa2005,Sun2005}, where
  the Coulomb interaction leads to strong correlations, e.g., at 150 meV in SrVO$_3$
  \cite{Yoshida2005}.
  Unexpectedly, kinks at 380 meV and 800 meV were reported recently for
  three different families of high-temperature superconductors at
  different doping levels \cite{Graf2006}.
  
  Interactions between electrons or their coupling to other degrees of
  freedom change the notion of $E_{\bm{k}}$ as the energy of an
  isolated electron or of a mode with infinite lifetime.  Namely, the
  interactions lead to a damping effect implying 
  that the dispersion relation is no longer a real function.  For systems with Coulomb 
  interaction FL theory predicts the existence of fermionic
  quasiparticles \cite{AGD}, i.e., exact one-particle states
  with momentum $\bm{k}$ and a real dispersion $E_{\bm{k}}$, at the Fermi surface
  and at zero temperature.  This concept can be extended to $\bm{k}$
  states sufficiently close to the Fermi surface (\emph{low-energy
    regime}) and at low enough temperatures, in which case the
  lifetime is now finite but still long enough for quasiparticles to
  be used as a concept.

  Outside the FL regime the notion of dispersive quasiparticles is, in
  principle, inapplicable since the lifetime of excitations is too
  short.  However, it is an experimental fact that $\bm{k}$-resolved
  one-particle spectral functions measured by ARPES often show
  distinct peaks also at energies far away from the Fermi
  surface.  The positions of those peaks change with $\bm{k}$, which
  means that the corresponding one-particle excitations are
  \emph{dispersive}, in spite of their rather short lifetime.
  It turns out that kinks in the dispersion relation are found in this energy
  region outside the FL regime
  \cite{Lanzara2002,Shen2002,He2001,Hwang2004,Hengsberger1999,Valla1999,Rotenberg2000,Higashiguchi2005,Schafer2004,Menzel2005,Yang2005,Aiura2004,Iwasawa2005,Sun2005,Ronning2003,Yoshida2005,Graf2006}.
  
  \begin{figure*}[t]
    \begin{center}\includegraphics[clip,height=90mm]{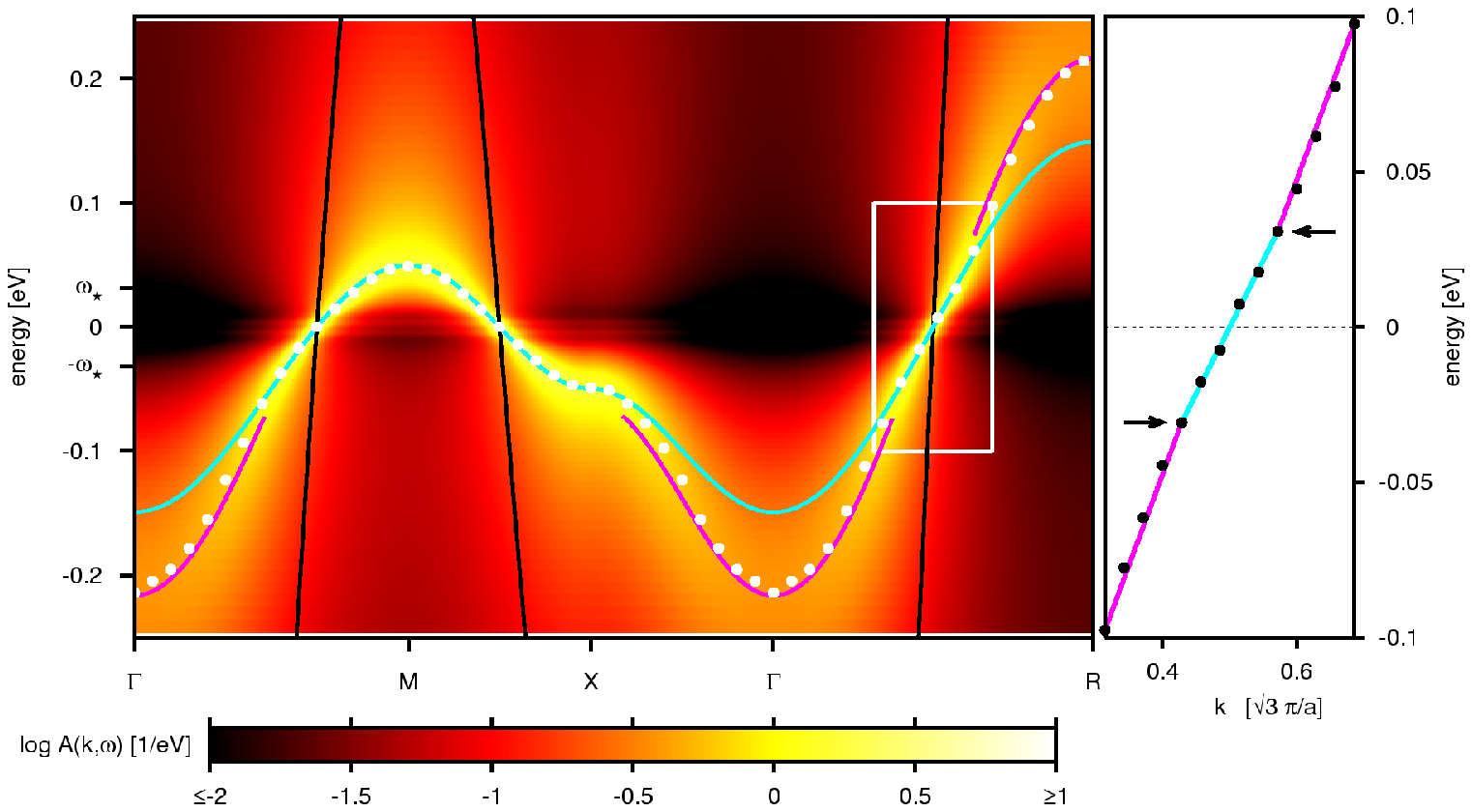}\vspace*{-5mm}\end{center}
    \caption{\mycaptionone%
      \vspace*{2mm}%
    }
    \label{fig:strong}
  \end{figure*}

  We describe here a novel mechanism leading to kinks in the
  dispersion of strongly correlated electrons, which does not require any coupling to
  phonons or other excitations.  To set the stage for this mechanism,
    which requires strong correlations, we consider first a weakly correlated
  system.  Imagine we inject an electron into the partially filled
  band at an energy close to the Fermi surface.  In this process the
  entire system becomes excited, leading to the generation of many
  quasiparticles and 
  -holes.  In view of their long lifetime the Coulomb interaction with
  other quasiparticles or -holes modifies their
  dispersion which, according to FL theory,
  becomes $E_{\bf k}=Z_{\text{FL}}\epsilon_{\bm{k}}$. Here
  $Z_{\text{FL}}$ is a FL renormalization factor and
  $\epsilon_{\bm{k}}$ is the bare (noninteracting) dispersion.  By
  contrast, an electron injected at an energy far from the Fermi level
  leads to excitations with only a short lifetime; their dispersion is
  hardly affected by the weak interaction,
  i.e.,  $E_{\bm{k}}\approx\epsilon_{\bm{k}}$ (see \myappendixname).  The
  crossover from the FL dispersion to the noninteracting dispersion
  can lead to kinks near the band edges which mark the termination
  point of the FL regime.  However, for weakly correlated metals
  ($Z_{\text{FL}}\lesssim1$) the slope of $E_{\bm{k}}$ changes only a
  little; hence the kinks are not very pronounced.
  
  The situation is very different in strongly correlated metals where
  $Z_{\text{FL}}$ can be quite small such that kinks can be
  well-pronounced.  The strong interaction produces a strong
  redistribution of the spectral weight in the one-particle spectral
  function.  Namely, the conducting band develops so-called Hubbard
  subbands, 
  whose positions are determined by the atomic energies.  For metallic
  systems a resonant central peak emerges around the Fermi level which
  lies between these subbands.  The central peak of this so-called
  three-peak structure is often interpreted as a ``quasiparticle
  peak'', but it will be shown below that genuine FL quasiparticles
  exist only in a \emph{narrow} energy range around the Fermi level.
  Outside this FL regime, but still inside the central peak, we
  identify a new \emph{intermediate-energy} regime, where the
  dispersion is given by $E_{\bf k}\approx Z_{\text{CP}}\epsilon_{\bm{k}}$.
  Here $Z_{\text{CP}}$ is a new renormalization factor, given by the
  weight of the \emph{c}entral \emph{p}eak, which differs significantly from
  $Z_{\text{FL}}$.  At these intermediate energies, which are much
  smaller than the interaction strength, an injected electron or hole
  is still substantially affected by the other electrons in the system.
  Therefore its dispersion is neither that of a free system, nor that
  of the (strongly renormalized) FL regime, but rather corresponds to  
  a \emph{moderately} correlated system
  ($Z_{\text{FL}}<Z_{\text{CP}}<1$).  As a consequence there occurs a
  crossover at an intermediate energy $\pm\omega_\star$ inside the
  central peak from $Z_{\text{FL}}$ renormalization to
  $Z_{\text{CP}}$ renormalization, which is visible as kinks in the
  dispersion.  As shown below, in a microscopic theory the position of those kinks and
  the breakdown of the FL regime are directly related.
  We emphasize that this mechanism yields kinks but does not
  involve coupling of electrons and collective modes; only strong
  correlations between electrons are required.

\begin{figure}[t]
    \begin{center}\includegraphics[clip,width=86mm]{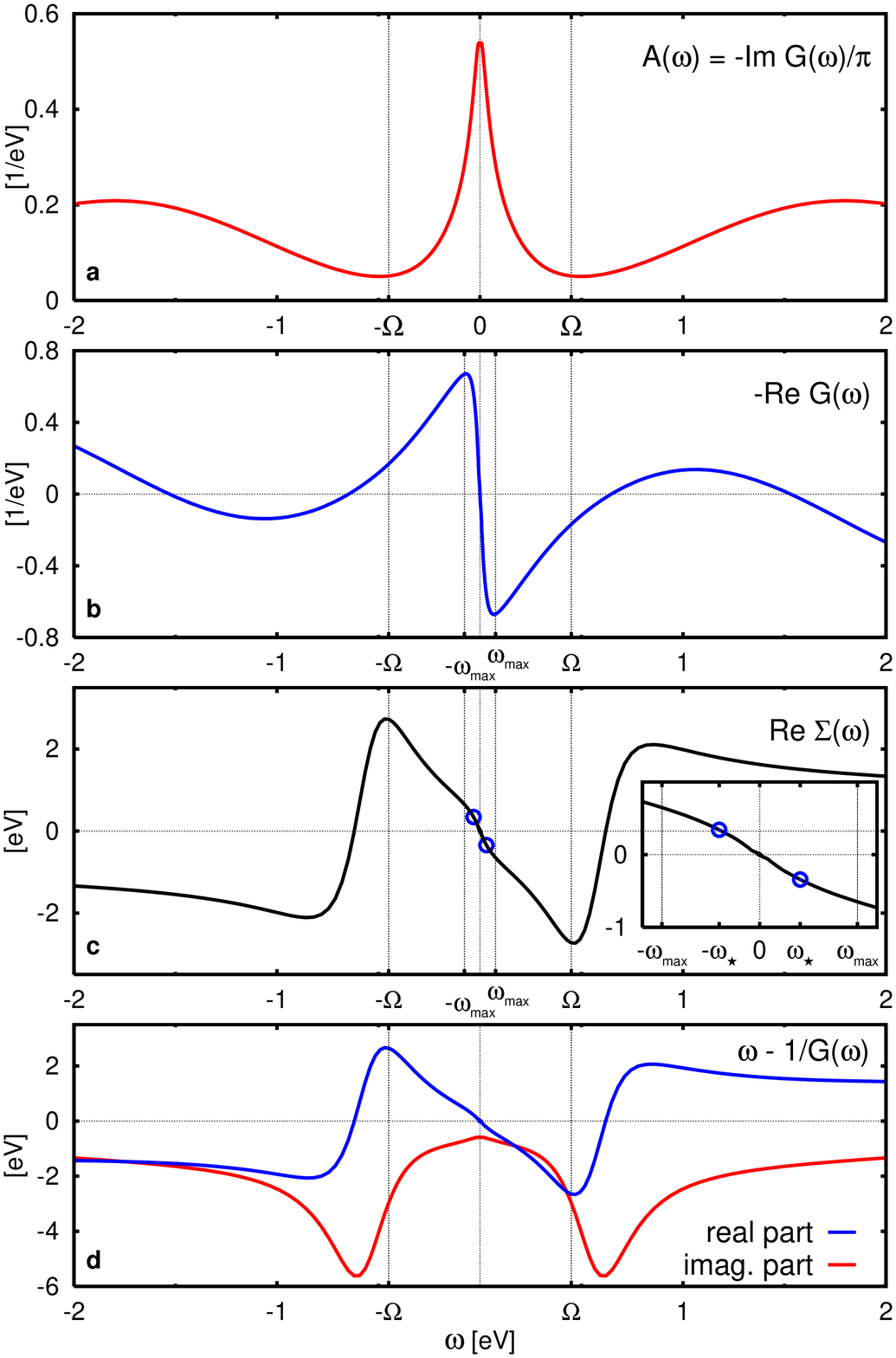}\vspace*{-5mm}\end{center}
    \caption{\mycaptiontwo%
      \vspace*{-5mm}%
    }
    \label{fig:kinks}
  \end{figure}

  For a microscopic description of these electronic kinks we use the
  Hubbard model, which is the generic model for strongly correlated
  electrons, and solve it by many-body dynamical mean-field theory
  (DMFT) \cite{Metzner89,Pruschke95,Georges96,PT},
  using the numerical renormalization group as an impurity solver.
  We focus on a single band with
  particle-hole symmetry and discuss the asymmetric case in the \myappendixname.
  For the strongly correlated Hubbard model (interaction $U$ $\approx$
  bandwidth) the
  dispersion relation is shown in Fig.~\ref{fig:strong} and the
  spectral function in Fig.~\ref{fig:kinks}a.
  The dispersion
  relation $E_{\bm{k}}$ crosses over from the Fermi-liquid regime (blue line in Fig.~\ref{fig:strong}) to the
  intermediate-energy regime (pink line in Fig.~\ref{fig:strong}), as described above, and shows pronounced
  kinks at the energy scale $\omega_\star=0.03$ eV. In some directions
  in the Brillouin zone these kinks may be less visible because the band structure is flat
  (e.g., near the X point in Fig.~\ref{fig:strong}).
  The behavior of $E_{\bm{k}}$ is now analyzed quantitatively.

  The physical quantity describing properties of one-particle
  excitations in a many-body system is the Green function or
  ``propagator''
  $G({\bm{k}},\omega)=(\omega+\mu-\epsilon_{\bm{k}}-\Sigma({\bm{k}},\omega))^{-1}$,
  which characterizes the propagation of an electron in the solid \cite{AGD}.  Here
  $\omega$ is the frequency, $\mu$ the chemical potential,
  $\epsilon_{\bm{k}}$ the bare dispersion relation, and
  $\Sigma({\bm{k}},\omega)$ is the self-energy, a generally complex
  quantity describing the influence of interactions on the propagation
  of the one-particle excitation, which vanishes in a noninteracting
  system.  The effective dispersion relation $E_{\bm{k}}$ of the
  one-particle excitation is determined by the singularities of
  $G({\bm{k}},\omega)$, which give rise to peaks in the spectral
  function $A({\bm{k}},\omega)=-\text{Im}G({\bm{k}},\omega)/\pi$.  If
  the damping given by the imaginary part of $\Sigma({\bm{k}},\omega)$
  is not too large, the effective dispersion is thus determined by
  $E_{\bm{k}}+\mu-\epsilon_{\bm{k}}-\text{Re}\Sigma({\bm{k}},E_{\bm{k}})=0$.
  Any kinks in $E_{\bm{k}}$ that do not originate from
  $\epsilon_{\bm{k}}$ must therefore be due to slope changes in
  $\text{Re}\Sigma({\bm{k}},\omega)$.

  In many three-dimensional physical systems the $\bm{k}$ dependence
  of the self-energy is less important than the $\omega$ dependence and can be neglected to a good approximation.  Then
  one may use the DMFT self-consistency equations to express
  $\Sigma({\bm{k}},\omega)=\Sigma(\omega)$ as
  $\Sigma(\omega)=\omega+\mu-1/G(\omega)-\Delta(G(\omega))$, where
  $G(\omega)=\int G({\bm{k}},\omega)\,d{\bm{k}}$ is the local Green
  function (averaged over $\bm{k}$) and $\Delta(G)$ is an
  energy-dependent hybridization function, expressed here as a
  function of $G(\omega)$ \cite{hybrid}.  The hybridization function
  describes how the electron at a given lattice site is
  quantum-mechanically coupled to the other sites in the system.  It
  plays the role of a dynamical mean-field parameter and its
  behavior is strongly dependent on the electronic correlations in the system.
  Fig.~\ref{fig:kinks}a shows a typical result for the
  integrated spectral function $A(\omega)=-\text{Im}G(\omega)/\pi$
  with the aforementioned three-peak structure. The corresponding
  real parts of the local propagator $G(\omega)$ and self-energy
  $\Sigma(\omega)$ are shown in Fig.~\ref{fig:kinks}b and
  Fig.~\ref{fig:kinks}c, respectively.

  Kinks in $\text{Re}\Sigma(\omega)$ appear at a new small
  energy scale which emerges quite generally for a three-peak spectral
  function $A(\omega)$.  Kramers-Kronig relations imply that
  $\text{Re}[G(\omega)]$ is small near the dips of $A(\omega)$,
  located at $\pm\Omega$. Therefore $\text{Re}[G(\omega)]$ has a
  maximum and a minimum at $\pm\omega_{\text{max}}$ \emph{inside the
    central spectral peak} (Fig.~\ref{fig:kinks}b). This directly leads
  to kinks in 
  $\text{Re}\Sigma(\omega)$ for the following reason.
  There are two contributions to $\Sigma(\omega)$:
  $\omega+\mu-1/G(\omega)$ and $-\Delta(G(\omega))$.  While
  $\text{Re}[\omega+\mu-1/G(\omega)]$ is \emph{linear} in the large
  energy window $|\omega|<\Omega$ (Fig.~\ref{fig:kinks}d), the
  term $-\text{Re}[\Delta(G(\omega))]$ is approximately proportional
  to $-\text{Re}[G(\omega)]$ (at least to first order in a moment
  expansion), and thus remains linear only in a much narrower energy
  window $|\omega|<\omega_{\text{max}}$.  The sum of
  these two contributions produces pronounced kinks in the real part
  of the self-energy at $\pm\omega_\star$, where
  $\omega_\star=(\sqrt{2}-1)\omega_{\text{max}}$ is the energy where
  $\text{Re}[G(\omega)]$ has maximum curvature (marked by blue
  circles in Fig.~\ref{fig:kinks}c).  The Fermi-liquid regime with
  slope $\partial \text{Re}\Sigma(\omega)/\partial\omega =
  1-1/Z_{\text{FL}}$ thus extends only throughout a small part of the
  central peak ($|\omega |<\omega_\star$). At intermediate
  energies ($\omega_\star<|\omega|<\Omega$) the slope
  is then given by $\partial\text{Re}\Sigma(\omega)/\partial\omega =
  1-1/Z_{\text{CP}}$. The kinks at $\pm\omega_\star$ mark the
  crossover between these two slopes.
  As a consequence there is also a kink at $\omega_\star$ in the effective band structure
  $E_{\bm{k}}$.

  The above analysis also explains why outside the FL regime $E_{\bm{k}}$ still follows the uncorrelated
  dispersion, albeit with a different renormalization $Z_{\text{CP}}$
  and a small offset $c$. This behavior is due to
  $\omega+\mu-1/G(\omega)$, the main contribution to the self-energy
  inside the central peak for $\omega_\star<|\omega|<\Omega$.
  In particular our analysis explains the dependence of
  $E_{\bm{k}}$ on $\bm{k}$ that was observed in previous DMFT studies of
  SrVO$_3$ \cite{Nekrasov06} (see \myappendixname).

  The FL regime terminates at the kink energy scale $\omega_\star$, which
  cannot be determined within FL theory itself.  The
  quantities $\omega_\star$, $Z_{\text{CP}}$, and $c$ can nevertheless
  all be expressed in terms of $Z_{\text{FL}}$ and the bare density of states
  alone;
  explicitly, one finds $\omega_\star=Z_{\text{FL}}(\sqrt{2}-1)D$, where $D$
    is an energy scale of the noninteracting system, e.g., $D$ is approximately given by
  half the bandwidth (see \myappendixname for details).  For weak correlations ($Z_{\text{FL}}\lesssim1$)
  the kinks in $E_{\bm{k}}$ thus merge with the band edges and are
  almost undetectable, as discussed above.  On the other hand, for
  increasingly stronger correlations ($Z_{\text{FL}}\ll 1$) the kinks
  at $\omega_\star/D\propto Z_{\text{FL}}$ move closer to the Fermi
  energy and deeper inside the central peak, whose width diminishes
  only as $\Omega/D\propto\sqrt{Z_{\text{FL}}}$ \cite{Bulla99}.

  The energy scale $\omega_\star$ involves only the bare band
  structure which can be obtained, for example, from band structure calculations, and the
  FL renormalization 
  $Z_{\text{FL}}=1/(1-\partial
  \text{Re}\Sigma(0)/\partial\omega)\equiv m/m^*$ known from, e.g.,
  specific heat measurements or many-body calculations.
  We note that since phonons are not involved in this mechanism,
  $\omega_\star$ shows no isotope effect.  For strongly interacting
  systems, in particular close to a metal-insulator transition
  \cite{PT}, $\omega_\star$ can become quite small, e.g., smaller
  than the Debye energy.
  
  An analysis similar to the one presented above also holds for
  systems with strong hybridization such as the high-temperature
  superconductors, where the overlap between $d$ and
  oxygen $p$ states 
  is important \cite{local}.  The assumption of a $\bm{k}$-independent self-energy
  may also be relaxed: if a correlation-induced three-peak spectral
  function $A({\bm{k}},\omega)$ is present for a certain range of
  momenta $\bm{k}$, the corresponding self-energies
  $\Sigma({\bm{k}},\omega)$ and effective dispersion $E_{\bm{k}}$ will
  also develop kinks, as can be proved using cluster extensions to
  DMFT.

  In conclusion, we showed that kinks in the electronic dispersion
  result from strong correlations alone, i.e., even without coupling
  to collective modes such as phonons. The energy of these kinks is a
  quantitative measure of electronic correlations in many-body
  systems; it marks the termination point of the Fermi liquid regime and
  can be as high as several hundred meV. These kinks are a
  fingerprint of a strongly correlated metal and are expected to be
  observable in many materials, including high-temperature
  superconductors.

  \myacknowledgements

  \makeatletter\galley@sw{}{\twocolumngrid}\makeatother
  
  \appendix

  \section*{Appendix: Technical details}

  \subsection*{Existence of kinks in the self-energy}

  A sufficient condition for kinks in the real part of the self-energy
  $\Sigma(\omega)$ and in the effective dispersion $E_{\bm{k}}$ is the existence of
  a correlation-induced three-peak structure in the
  spectral function $A(\omega)=-\text{Im}G(\omega)/\pi$.  These kinks are located
  at energies inside the central peak of $A(\omega)$.  This
  can be derived from the DMFT self-consistency condition,
  \begin{align}
    \Sigma(\omega)=\omega+\mu-1/G(\omega)-\Delta(G(\omega))
    \,,\label{eq:sigmacontributions}
  \end{align}
  where $G(\omega)$ is the local Green function and $\Delta(G)$ is the
  hybridization function \cite{hybrid}.

  Suppose that $A(\omega)$ has three well-developed peaks with dips at
  $\Omega_\pm$ (Fig.~\ref{fig:kinks}a), i.e., the central peak is
  located in the interval $\Omega_-<\omega<\Omega_+$ (allowing
  for a general, asymmetric case).  Kramers-Kronig relations imply
  that $\text{Re}G(\omega)$ becomes small in the vicinity of
  $\Omega_\pm$ and thus has extrema inside the central spectral peak
  (Fig.~\ref{fig:kinks}b).  Consider now the complex function
  $\omega-1/G(\omega)$.  Peaks and dips in $A(\omega)$ are reflected
  as dips and peaks in $\text{Im}[-1/G(\omega)]$, respectively (red
  line in Fig.~\ref{fig:kinks}d). By Kramers-Kronig relations, the
  peaks of $\text{Im}[-1/G(\omega)]$ near $\Omega_\pm$ imply zeros in
  $\omega-\text{Re}[1/G(\omega)]$, which thus has one maximum and one
  minimum inside the central peak (blue line in
  Fig.~\ref{fig:kinks}d). Hence we find that \emph{inside the central
    peak} $\text{Re}[\omega-1/G(\omega)]$ \emph{is monotonous and may
    be approximated by a straight line}, provided $A(\omega)$ is
  sufficiently smooth.  Thus there are two different contributions to
  $\text{Re}\Sigma(\omega)$ (Fig.~\ref{fig:kinks}c):
  (i) $\text{Re}[\omega+\mu-1/G(\omega)]$ is
  approximately linear in $\omega$ throughout the central spectral
  peak, while (ii) $\text{Re}[-\Delta(G(\omega))]$
  is linear in a smaller interval  $\omega_{\star,-}<\omega<\omega_{\star,+}$, thus
  \emph{leading to
    kinks} \emph{in} $\text{Re}\Sigma(\omega)$ \emph{inside the central
    spectral peak}.  Below we determine the location $\omega_{\star,\pm}$ of these kinks  
  and the resulting effective dispersion relation $E_{\bm{k}}$.

  \subsection*{Location of kinks in the self-energy}

  As discussed above, since $\text{Re}[\omega+\mu-1/G(\omega)]$ is
  approximately linear in this interval, we can expand the inverse of
  the local propagator as $1/G(\omega)=z_0+z_1\omega+O(\omega^2)$ for
  small $\omega$. We rewrite this as
  \begin{align}
    G(\omega)
    &=
    \frac{Z_{\text{CP}}}{\omega-\omega_0+i(\gamma+\gamma'\omega)} +
    O(\omega^2)
    \,,\label{eq:greenexpansion}
  \end{align}
  i.e., $z_0=(-\omega_0+i\gamma)/Z_{\text{CP}}$ and
  $z_1=(1+i\gamma')/Z_{\text{CP}}$.  For the particle-hole symmetric
  case $\omega_0$ and $\gamma'$ are zero.  We note that the in
  vicinity of $\omega=0$ the local Green function $G(\omega)$ can thus be
  approximated by a simple pole. When neglecting $\gamma'$, the
  parameter $Z_{\text{CP}}$ equals the weight of the central peak of
  $A(\omega)$, which is approximated by a Lorentzian.

  The parameters $Z_{\text{CP}}$, $\gamma$, $\omega_0$, $\gamma'$ are
  determined as follows. We employ the DMFT self-consistency equation
  \begin{align*}
    G(\omega)
    &=
    G_0(\omega+\mu-\Sigma(\omega))
    \,,
  \end{align*}
  which is equivalent to Eq. (\ref{eq:sigmacontributions})
  \cite{hybrid}. Here
  \begin{align*}
    G_0(z)=\int d\epsilon \,\frac{\rho_0(\epsilon)}{z-\epsilon+i0}
  \end{align*}
  is the propagator and $\rho_0(\omega)$ the density of states (DOS) for
  the noninteracting case.  We also have the Fermi-liquid relations
  $\partial\Sigma/\partial\omega|_{\omega=0}=1-1/Z_{\text{FL}}$ and
  Luttinger's theorem
  \mymhcite,
  which reduces to $\mu-\Sigma(0)=\mu_0$ in
  DMFT, where $\mu_0 $ is the chemical potential for the corresponding
  noninteracting system.  Altogether this leads us to the equations
  \begin{align*}
    z_0
    &=
    \frac{1}{G_0(\mu_0)}
    \,,
    &
    z_1
    &=
    \frac{-G_0'(\mu_0)}{Z_{\text{FL}}G_0(\mu_0)^{2}}
    \,,  
  \end{align*}
  which are immediately solved by taking real and imaginary parts, i.e.,
  \begin{align*}
    Z_{\text{CP}}
    &=
    \frac{1}{\text{Re}\,z_1}
    \,,
    &
    \gamma
    &=
    \frac{\text{Im}\,z_0}{\text{Re}\,z_1}
    \,,
    \\
    \omega_0
    &=
    \frac{-\text{Re}\,z_0}{\text{Re}\,z_1}
    \,,
    &
    \gamma'
    &=
    \frac{\text{Im}\,z_0}{\text{Re}\,z_1}
    \,.
  \end{align*}
  The parameters $Z_{\text{CP}}$, $\gamma$, $\omega_0$, $\gamma'$ are
  thus determined by $Z_{\text{FL}}$ and the bare DOS alone.  For the
  parameters in Fig.~\ref{fig:strong} we obtain $Z_{\text{CP}}=0.135$,
  $\gamma=0.076$ eV, $\omega_0=0$, $\gamma'=0$.

  Using the expansion (\ref{eq:greenexpansion}), the first
  contribution to the self-energy [Eq.~(\ref{eq:sigmacontributions})]
  becomes
  \begin{align}
    \text{Re}[\omega+\mu-1/G(\omega)]=\text{const}+(1-1/Z_{\text{CP}})\omega
    \,,\label{eq:contribution1}
  \end{align}
  i.e., this function is linear inside the central peak. On the other
  hand $\text{Re}[\Delta(G(\omega))]$ is linear only on the narrower scale
  $|\omega|<|\omega_{\star,\pm}|\ll|\Omega_{\pm}|$ and is thus
  responsible for kinks in $\text{Re}[\Sigma(\omega)]$ at
  $\omega_{\star,\pm}$, which are located inside the central peak.  This
  location can now be calculated by inserting the linear ansatz for
  $1/G$ into $\Delta(G)$.  To identify the relevant energy scales we
  proceed by expanding the DMFT self-consistency equation
  \cite{hybrid} as
  \begin{align*}
    \Delta(G)=(m_2-m_1^2)G+(m_3-3m_1m_2+2m_1^3)G^2+\cdots
    \,,
  \end{align*}
  where $m_i$ are the moments of the bare DOS.  This moment expansion
  terminates after
  the first term  for a semi-elliptical DOS;
  we omit the other terms in the following
  discussion.  The kinks are located roughly at the extrema of
  $\text{Re}[\Delta(G(\omega))]$, i.e., at
  \begin{align}
    \omega_{\text{max},\pm}
    =
    \omega_0\pm\frac{\gamma+\gamma'\omega_0}{\sqrt{1+\gamma'^2}}
    \,.\label{eq:omegamax}
  \end{align}
  To better understand the energy scales involved we assume
  particle-hole symmetry for the moment and use the first-order
  expansion of $\Delta(G)$. Then we find
  \begin{align*}
    \omega_{\text{max},\pm}
    \approx
    \pm\gamma\approx\pm2qZ_{\text{FL}}\sqrt{m_2-m_1^2}
    =:Z_{\text{FL}}D
    \,,
  \end{align*}
  with $q=(p+1/p)/2\geq1$ and $p=\pi \rho_0(\mu_0)\sqrt{m_2-m_1^2}$.
  \emph{The kink location is thus given by $Z_{\text{FL}}$ times a
    non-interacting energy scale $D$ which depends on the details of
    the DOS.} For example, for a half-filled band with semi-elliptical
  model DOS $D$ is given by half the bandwidth.
  
  An improved estimate of the kink energy scale $\omega_\star$ is
  obtained from the maximum curvature of
  $\text{Re}[-\Delta(G(\omega))]$. The relevant solutions of
  $\partial^3\text{Re}G/\partial\omega^3=0$ are
  \begin{align}
    \omega_{\star,\pm}=\omega_0\mp
    \frac{\gamma+\gamma'\omega_0}{\sqrt{1+\gamma'^2}}
    \left[1-\sqrt{2}\left(1\pm\frac{\gamma'}{\sqrt{1+\gamma'^2}}\right)^{1/2}\right]
    \,,\label{eq:omegastar}
  \end{align}
  which reduces to $\omega_\star=Z_{\text{FL}}(\sqrt{2}-1)D$ for the
  particle-hole symmetric case.  We obtain
  $\omega_\star=|\omega_{\star,\pm}|=0.03$ eV for the parameters of
  Fig.~\ref{fig:strong}. This agrees well with the location of the
  kinks in $\text{Re}\Sigma(\omega)$, as seen in
  Fig.~\ref{fig:kinks}c, where $\pm\omega_\star$ is marked by blue
  circles.

  \subsection*{Effective dispersion relation}
  
  For energies inside the central peak we now determine the effective
  dispersion $E_{\bm{k}}$, which is defined as the frequency $\omega$)
  where $A(\bm{k},\omega)$ has a maximum.  Neglecting the $\omega$
  dependence of $\text{Im}\Sigma(\omega)$, kinks in $E_{\bm{k}}$ occur
  at the same energy $\omega_\star$ as the kinks in
  $\text{Re}\Sigma(\omega)$.  We approximate $\text{Re}\Sigma(\omega)$
  by a piecewise linear function with slope $1-1/Z_{\text{FL}}$ inside
  the Fermi liquid regime and $1-1/Z_{\text{CP}}$ in the
  intermediate-energy regime, i.e.,
  \begin{align}
    \text{Re}[\Sigma(\omega)-\Sigma(0)]
    &\notag
    \\
    &\!\!\!\!\!\!\!\!\!\!\!\!\!\!\!\!\!\!\!\!\!\!\!\!\!\!\!\!\!\!\!\!\!\!\!\!\!\!\!\!\!\!
    =
    \left\{
      \begin{array}{@{}l@{}l@{}}
        a_-+(1-1/Z_{\text{CP}})\omega
        & \text{  for  }\Omega_{\star,-}<\omega<\omega_{\star,-}
        \\
        (1-1/Z_{\text{FL}})\omega
        & \text{  for  }\omega_{\star,-}<\omega<\omega_{\star,+}
        \\
        a_++(1-1/Z_{\text{CP}})\omega
        & \text{  for  }\omega_{\star,+}<\omega<\Omega_{\star,+}
      \end{array}
    \right.
    .\label{eq:sigmapiecewise}
  \end{align}
  This approximation assumes that the self-energy constribution
  (\ref{eq:contribution1}) dominates over
  $\text{Re}[-\Delta(G(\omega))]$ outside the Fermi-liquid regime.
  Here $a_{\pm}=-(1/Z_{\text{FL}}-1/Z_{\text{CP}})\omega_{\star,\pm}$
  is required for continuity.  We obtain the effective dispersion from
  the equation
  $E_{\bm{k}}+\mu-\epsilon_{\bm{k}}-\text{Re}\Sigma(E_{\bm{k}})=0$ and
  use the approximation (\ref{eq:sigmapiecewise}). This yields
  \begin{align}
    \!\!\!\!\!\!\!\!E_{\bm{k}}
    =
    \left\{
      \begin{array}{@{}l@{}l@{}}
        Z_{\text{CP}}(\epsilon_{\bm{k}}-\mu_0)+c_-
        & \text{  for  }\Omega_{\star,-}<E_{\bm{k}}<\omega_{\star,-}
        \\
        Z_{\text{FL}}(\epsilon_{\bm{k}}-\mu_0)
        & \text{  for  }\omega_{\star,-}<E_{\bm{k}}<\omega_{\star,+}
        \\
        Z_{\text{CP}}(\epsilon_{\bm{k}}-\mu_0)+c_+
        & \text{  for  }\omega_{\star,+}<E_{\bm{k}}<\Omega_{\star,+}
      \end{array}
    \right.
    .\label{eq:dispersionpiecewise}
  \end{align}
  The effective dispersion thus follows the bare dispersion with 
  two different renormalization factors: $Z_{\text{FL}}$ in the Fermi liquid regime
  and $Z_{\text{CP}}$ in the intermediate-energy regime. The offset in
  Eq. (\ref{eq:dispersionpiecewise}) is given by
  \begin{align*}
    c_\pm
    =
    Z_{\text{CP}}a_\pm
    =
    -\left(\frac{Z_{\text{CP}}}{Z_{\text{FL}}}-1\right)\omega_{\star,\pm}
    \,.
  \end{align*}
  In the particle-hole symmetric case (half-filled band with symmetric
  DOS and $\mu_0=0$) we find the effective dispersions
  $E_{\bm{k}}=Z_{\text{FL}}\epsilon_{\bm{k}}$ and
  $E_{\bm{k}}=Z_{\text{CP}}\epsilon_{\bm{k}}\mp c$ with $c=|c_\pm|$,
  respectively.  For the parameters in Fig.~\ref{fig:strong} we obtain
  $c=0.018$ eV. As Fig.~\ref{fig:strong} shows, the agreement of these
  renormalized dispersions with the observed maxima of $A(\omega)$ is
  very good.

  \subsection*{Further examples}

  We close with three figures in order to put the occurrence of
  electronic kinks into broader perspective.

\begin{figure*}[b]
    \begin{center}\vspace*{-2mm}\includegraphics[height=85mm]{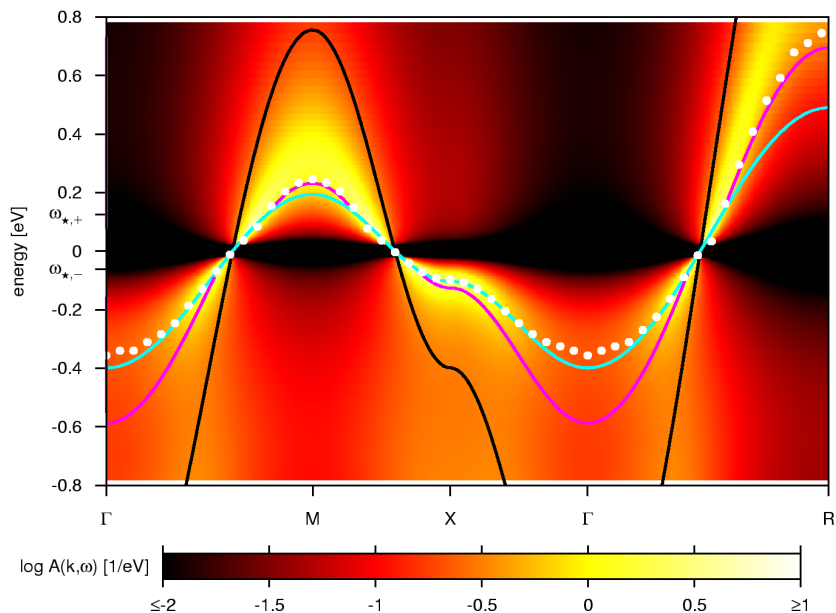}\vspace*{-5mm}\end{center}
    \caption{%
      Intensity plot of the spectral function $A(\bm{k},\omega)$ for a
      strongly correlated system without particle-hole symmetry
      (Hubbard model in DMFT, cubic lattice, $U$=4 eV, bandwidth
      $W=3.46$ eV, $n$=0.8, $Z_{\text{FL}}$=0.26).  The effective
      dispersion $E_{\bm{k}}$ (white dots) follows
      $Z_{\text{FL}}\epsilon_{\bm{k}}$ (blue line) near the Fermi
      level. At $\omega_{\star,+}$=0.062 eV it crosses over to
      $Z_{\text{CP}}\epsilon_{\bm{k}}+c_+$ (pink line at positive
      energies, $Z_{\text{CP}}=0.40$, $c_+$=0.035 eV).  The
      crossover between these two regimes leads to a pronounced kink.
      Here $\omega_\star$, $Z_{\text{CP}}$, and $c_+$ were calculated
      from $Z_{\text{FL}}$ and the uncorrelated band structure as
      described above.
      On the other hand, below the Fermi level there
      is no crossover to $Z_{\text{CP}}\epsilon_{\bm{k}}+c_-$ (pink
      line at negative energies) because the lower Hubbard subband is not
      separated well enough from the central spectral peak.%
       \vspace*{-3mm}%
    }
    \label{fig:offhalf}
  \end{figure*}

  A strongly correlated system \emph{without particle-hole} symmetry
  is shown in Fig.~\ref{fig:offhalf}.  For this less than half-filled
  band we find a pronounced kink in the effective dispersion
  above the Fermi level. On the other hand, there is no kink below the
  Fermi level because the lower Hubbard band in $A(\omega)$ is not
  separated well enough from the central spectral peak.

\begin{figure*}[p]
    \begin{center}\vspace*{-2mm}\includegraphics[height=85mm]{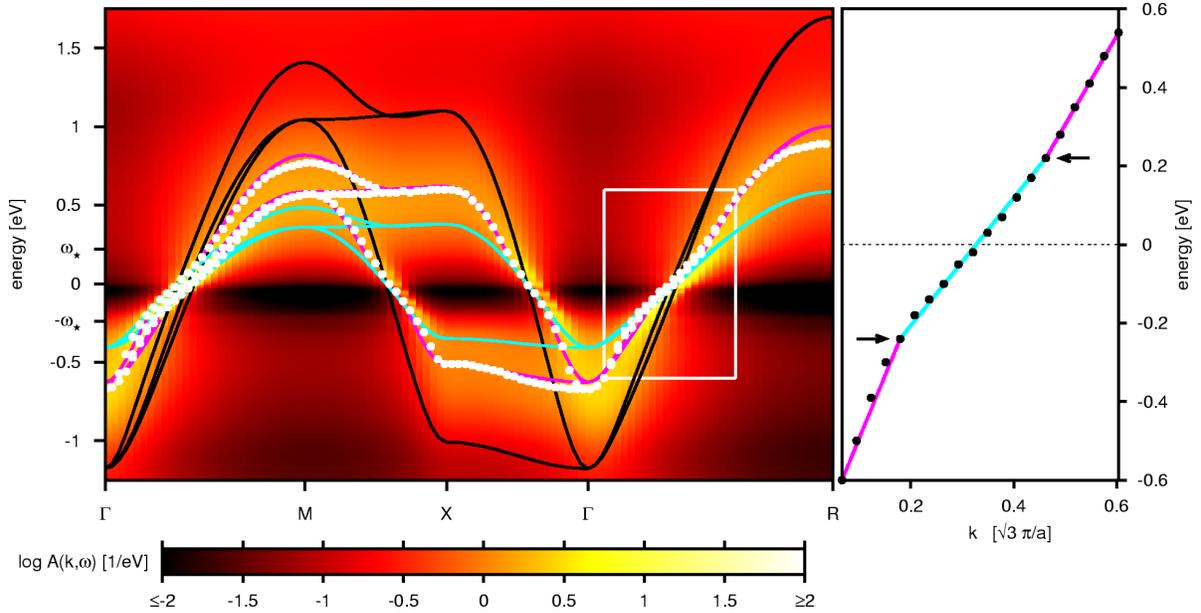}\vspace*{-4mm}\end{center}
    \caption{%
      Kinks in the dispersion relation $E_{n\bm{k}}$ (white dots)
      obtained for SrVO$_3$ with LDA+DMFT. In the vicinity of the
      Fermi energy it follows the LDA band structure
      $\epsilon_{n\bm{k}}$ (black lines) renormalized by a
      Fermi-liquid factor $Z_{\text{FL}}=0.35$, i.e.
      $E_{n\bm{k}}=Z_{\text{FL}}\epsilon_{n\bm{k}}$ (blue line).
      Outside the Fermi-liquid regime the dispersion relation follows
      the LDA band structure with a different renormalization,
      $E_{n\bm{k}}=Z_{\text{CP}}\epsilon_{n\bm{k}}+ c_\pm$ (pink
      line), with $Z_{\text{CP}}=0.64$, $c_+$=0.086 eV, $c_-=$0.13 eV
      (as determined from the linear
      approximation to $1/G$, in contrast to \cite{Nekrasov06}).
      Along the directions $\Gamma$-M and $\Gamma$-R the crossover
      between the two regimes leads to kinks at energies
      $\omega_{\star,+}$=0.22 eV and $\omega_{\star,-}$=-0.24 eV
      in the effective dispersion. These kinks are marked by arrows
      in the plot on the right, which corresponds to the white frame 
      and shows the approximately piecewise linear dispersion of the
      lowest-lying band.
      From the intensity plot of the spectral
      function $A(\bm{k},\omega)$ we note that in the
      intermediate-energy regime the resonance is rather broad but
      nonetheless dispersive. The LDA+DMFT calculation was performed
      for Hubbard interaction $U=5.55$ eV and exchange interaction
      $J=1.0$ eV \cite{Nekrasov06}; due to the degeneracy of the
      $t_{2g}$ band the self-energy obtained from DMFT is a diagonal
      matrix with equal elements. The results were obtained at
      temperature $T=0.1$ eV with QMC as the impurity solver.%
      \vspace*{-3mm}%
    }
    \label{fig:srvo3}
  \end{figure*}

  For a degenerate \emph{multi-band} system the analysis is very
  similar to that for a single-band system; results for
  SrVO$_3$ (with three degenerate correlated bands) 
  are given in Fig.~\ref{fig:srvo3}.

\begin{figure*}[p]
    \begin{center}\vspace*{-2mm}\includegraphics[height=85mm]{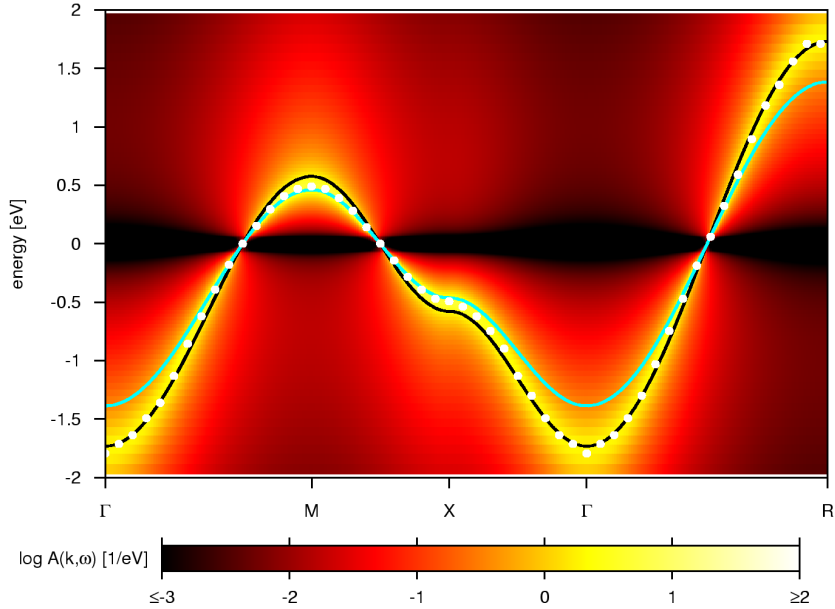}\vspace*{-5mm}\end{center}
    \caption{%
      Intensity plot of the spectral function $A(\bm{k},\omega)$ for a
      weakly correlated system (Hubbard model in DMFT, cubic lattice,
      $U$=1 eV, bandwidth $W=3.46$ eV, $n$=1, $Z_{\text{FL}}$=0.80).
      The effective dispersion $E_{\bm{k}}$ (white dots) follows
      $Z_{\text{FL}}\epsilon_{\bm{k}}$ (blue line) near the Fermi
      level and crosses over to bare dispersion $\epsilon_{\bm{k}}$
      (black line) at higher energies. The crossover between these two
      regimes does not lead to a sharp kink.
      As in Fig.~\ref{fig:strong}, the Gaussian DOS
      for the hypercubic lattice with
      $t_*$=1 eV was used, which for a three-dimensional cubic lattice
      corresponds to $t=t^*/\sqrt{6}$, i.e., bandwidth $W=3.46$ eV.%
      \vspace*{-3mm}%
    }
    \label{fig:weak}
  \end{figure*}

  Finally, Fig.~\ref{fig:weak} shows results for a \emph{weakly
    correlated system} with particle-hole symmetry.  For weakly
  correlated systems the spectral function $A(\omega)$ usually has a
  single peak, given by the bare DOS with additional
  broadening. The real part of the self-energy is much smaller than in
  the strongly correlated case and typically has a broad maximum near
  the lower and a minimum near the upper band edge.  At these extrema
  the effective dispersion $E_{\bm{k}}$ crosses over from the FL
  dispersion $Z_{\text{FL}}\epsilon_{\bm{k}}$ to the free dispersion
  $\epsilon_{\bm{k}}$, but the corresponding kinks are very faint
  since $Z_{\text{FL}}$ is close to 1. Since
  $\text{Re}[\omega-1/G(\omega)]$ is approximately linear throughout
  the band, these weak kinks are located at the extrema of
  $\text{Re}[-\Delta(G(\omega))]\approx-(m_2-m_1^2)\text{Re}[G(\omega)]$,
  i.e., at $\omega_{\text{max},\pm}$ [Eq.~(\ref{eq:omegamax})]. For the
  parameters of Fig.~\ref{fig:weak} we calculate
  $\omega_{\text{max}}$=0.71 eV, which agrees well with the observed
  location of the crossover.

  \mybibliography
  \onecolumngrid
 
\begin{thebibliography}{99}
\bibitem{hybrid} In DMFT, $\Delta(G)$ is determined by the requirement
    $G(\omega)=G_0(\omega+\mu-\Sigma(\omega))$, i.e.,
    $G_0(\Delta(G)+1/G)=G$.  Here $G_0(\omega)$ is the local Green
    function in the absence of interactions.
\bibitem{local} The mechanism discussed here applies to systems with
    localized d or f-orbitals, where the local interaction between
    electrons is dominant. The long-range part of the interaction is
    viewed on a mean-field level as in a local-density approximation
    theory for band structure calculations.
\end{thebibliography}
\end{document}